\documentstyle[prl,aps,epsf]{revtex}
\draft
\epsfclipon

\begin{document}

\twocolumn[\hsize\textwidth\columnwidth\hsize\csname 
@twocolumnfalse\endcsname
\title{High-field side of Superconductor-Insulator Transition}

\author{Tatyana I. Baturina$^1$,
D.~R.~Islamov$^1$, Z.~D.~Kvon$^1$, M.~R.~Baklanov$^2$, A.~Satta$^2$}

\address{$^1$Institute of Semiconductor Physics, 13 Lavrentjev Ave.,
630090 Novosibirsk, Russia\\
$^2$IMEC, Kapeldreef 75, B-3001 Leuven, Belgium}

\maketitle

\begin{abstract}
We report the experimental observation of a magnetic-field-tuned
superconductor-insulator transition (SIT) in ultrathin TiN films.
The low temperature transport properties of these films show
scaling behavior consistent with a transition driven by quantum
phase fluctuations in two-dimensional superconductor. The
magnetoresistance reveals peak and a subsequent decrease in fields
higher than the critical field. The temperature dependences of the
isomagnetic resistance data on the high-field side of the SIT have
been analyzed and the transition from insulating to metallic phase
is found, with at high fields the zero-temperature asymptotic
value of the resistance being equal to $h/e^2$.
\end{abstract}


]
\bigskip
\narrowtext

\section{Introduction}

The zero-temperature superconductor-insulator transition (SIT) is
a phase transition of the second order and is driven by
fluctuations of a quantum nature. The superconducting phase is
considered to be a condensate  of Cooper pairs with localized
vortices, and the insulating phase is a condensate of vortices
with localized Cooper pairs. The theoretical description based on
this assumption was suggested in Ref.~\cite{Fisher}. According to
the theory: (i)~the film resistance $R$ near the field-induced SIT
at low temperature $T$ in the vicinity of the critical field $B_c$
is a function of one scaling variable $\delta=(B-B_c)/T^{1/\nu
z}$, with $\nu z$ being a parameter; (ii)~at the transition point
the film resistance has the universal value $h/4e^2$ (the quantum
resistance for Cooper pairs). Although much work has been done,
and in many systems the scaling relations hold, the SIT in
disordered films remains a controversial subject, especially
concerning the insulating phase and the bosonic conduction on the
high-field side of the SIT at $B>B_c$. In some works the behavior
of the resistance in this region can be explained by
weak-localization theory. In Ref.~\cite{Destr} it has been found
that, on the high-field side of the transition, the
magnetoresistance reaches a maximum and the phase can be
insulating as well as metallic, with at high fields the
zero-temperature asymptotic value of the resistance is
approximately equal to $R_c$ (the value of the resistance at
$B=B_c$). Here we present a careful examination of the presence of
the insulating phase and its evolution on the high-field side of
the SIT.

\section{Experiments and discussion}

A TiN film with a thickness of 5~nm was formed on 100~nm of
SiO$_2$ grown on top of $<$100$>$ Si substrate by atomic layer
chemical vapour deposition at 400$^\circ$C \cite{ALCVD}. Analysis
by atomic force spectroscopy shows that the film exhibits low
surface roughness and consists of a dense packing of grains, with
a rather narrow distribution of grain size and the average size is
roughly $\sim 30$~nm.

The samples for the transport measurements were fabricated into
standard Hall bridges using conventional UV lithography and
subsequent plasma etching. The channel length and width were 250
and 50~$\mu$m, respectively. The magnetoresistance measurements
were performed in a temperature-stabilized dilution refrigerator.
The magnetic field was applied perpendicular to the film. Four
terminal transport measurements were performed using standard low
frequency techniques. The resistance data were taken at a
measurement frequency of 10~Hz with an ac current $1$~nA.

\begin{figure}
\centerline{\epsfxsize3in\epsfbox{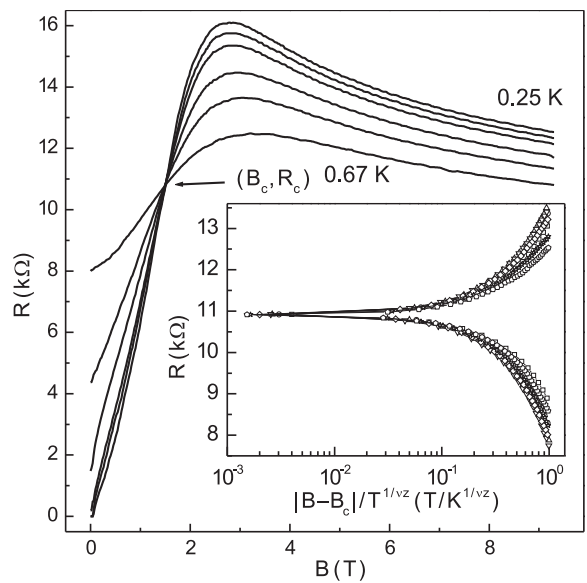}\bigskip}
\caption{Low-temperature isotherms in the $(B,R)$ plane.
Different curves represent different temperatures: 0.25, 0.38,
0.42, 0.51, 0.61, and 0.67 K. The point of intersection:
$B_{c}=1.52$~T is the critical magnetic field, and
$R_{c}=10.9$~k$\Omega$ is the critical resistance. The inset shows
a scaled plot of the same data with $\nu z=1$.}
\label{fig1}
\end{figure}

\begin{figure}
\centerline{\epsfxsize3in\epsfbox{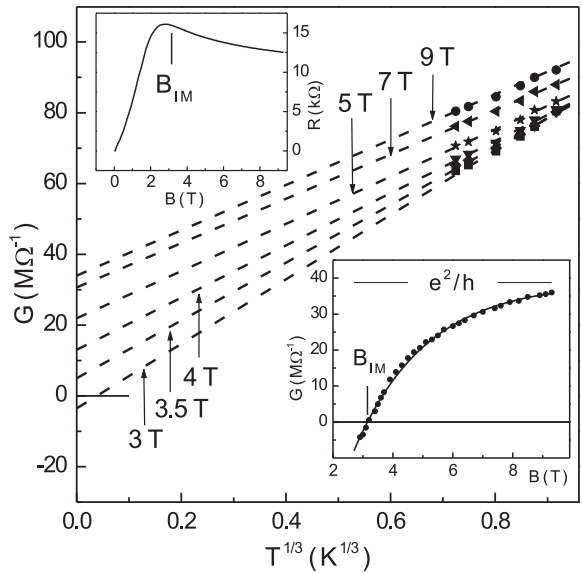}\bigskip}
\caption{Conductance $G = 1/R$ vs $T^{1/3}$ at different magnetic
fields on the high-field side of the SIT. The lower inset shows
the magnetic field dependence of the zero temperature conductance
determined from extrapolations in accordance with Eq. (1)
(symbols). The upper inset: $B_{IM}$ marks the value of the field
of the insulator-metal transition.}
\label{fig2}
\end{figure}

A typical set of  resistance per square vs magnetic field data is
given in Fig. 1. The main features of these results are the
presence of  the point of intersection ($B_c, R_c$) and the
negative magnetoresistance in high fields. Using the $B_c$, we
plot the same data against the scaling variable $|B-B_c|/T^{1/ \nu
z}$, and adjust the power $\nu z$ to obtain the best visual
collapse of the data. Such behavior, previously regarded as the
main evidence of the existence of SIT,  is actually not
incontestable proof of the presence of the insulating phase at
$B>B_c$. Following the approach of the authors in~\cite{Destr}, we
analyze the temperature dependence of the film resistance on the
high-field side of the SIT in terms of 3D ``bad'' metals in the
vicinity of the metal-insulator transition \cite{ReviewAA}:
\begin{equation}
G (T)=a+bT^{1/3}.
\label{equAA}
\end{equation}
The sign of the parameter $a$ discriminates between a metal and an
insulator at $T \to 0$. If $a>0$, it yields the zero temperature
conductance $G(0) \equiv a$, whereas negative $a$ points to
activated conductance at lower temperatures. The  temperature
dependence of the high-field conductance at different magnetic
fields is shown in Fig.~2. It is well described by Eq.~(1). As
$G(0)$ is negative in fields higher than the critical field we can
conclude that this phase is insulating. With further increase of
$B$ the resistance decreases. This behavior is in agreement with
the concept of localized Cooper pairs in the insulating phase, and
that of field-induced Cooper-pair breaking. Extrapolation to $T=0$
allows us not only to determine the field of the insulator-metal
transition ($B_{IM}$), but also the magnetic field dependence of
the zero temperature conductance. The result of this procedure is
presented in the lower inset to Fig.~2. The zero temperature
conductance at $B>B_{IM}$ is well described by an empirical
expression $G(B)=e^2/h(1-\exp [(B_{IM}-B)/B^*])$ shown by the
solid line in the lower inset to Fig.~2. In contrast to the
statement of the authors in~\cite{Destr}, the resistance at $T \to
0$ on the high-field side of the SIT approaches the value of
$h/e^2$ rather than $R_c$. The exponential dependence of
$G(T=0,B)$ may result from a broad dispersion of the binding
energies of localized Cooper pairs.

%\begin{ack}
This work was supported by Grants RFBR 00-02-17965, RFBR
02-02-16782 and by the programs ``Low-dimensional and Mesoscopic
Condensed Systems'' of the Russian Ministry of Science, Industry,
and Technology, and ``Quantum Macrophysics'' of the Russian
Academy of Science.

%\end{ack}

\end{document}